\documentclass[10pt,prd,twocolumn,floatfix,tightenlines,showpacs,showkeys,preprintnumbers,altaffilletter, nofootinbib, superscriptaddress, longbibliography]{revtex4-1}
\usepackage{graphicx,mathtools,tabu}
\usepackage{epstopdf}
\usepackage{amsmath}
\usepackage{amsfonts}
\usepackage[colorlinks=true,citecolor=red,linkcolor=blue,breaklinks=true]{hyperref}
\usepackage[dvipsnames]{xcolor}
\usepackage{accents}
\usepackage[figure, table]{hypcap}
\usepackage{amssymb}
\usepackage{appendix}
\usepackage{comment}
\usepackage{bbold}
\usepackage{color}
\usepackage{slashed}
\usepackage{subcaption}
\usepackage{setspace}
\usepackage{footnote}
\usepackage{multirow}
\usepackage{braket}
\usepackage[normalem]{ulem}
\usepackage[utf8]{inputenc}
\usepackage{mathrsfs}
\usepackage{enumitem}
\usepackage{microtype}
\usepackage{orcidlink}
\begin{document}
\preprint{}
\title{}
\title{Machine Learning Detection of Non-Axisymmetric Fast Flavor Instabilities in Compact Objects}

\author{Madhurima Chakraborty\orcidlink{0000-0002-9960-8467}}
\email{madhurima@as.edu.tw}
\affiliation{Institute of Physics, Academia Sinica, Taipei 115, Taiwan}

\author{Sajad Abbar\orcidlink{0000-0001-8276-997X}}
\email{abbar@mpp.mpg.de}
\affiliation{Max-Planck-Institut {f\"ur} Physik (Werner-Heisenberg-Institut), Boltzmannstr. 8, 85748 Garching, Germany}

\author{Meng-Ru Wu\orcidlink{0000-0003-4960-8706}}
\email{mwu@as.edu.tw}
\affiliation{Institute of Physics, Academia Sinica, Taipei 115, Taiwan}
\affiliation{Institute of Astronomy and Astrophysics, Academia Sinica, Taipei 106, Taiwan}
\affiliation{Physics Division, National Center for Theoretical Sciences, Taipei 106, Taiwan}

\author{Francois Foucart\orcidlink{0000-0003-4617-4738}}
\email{francois.foucart@unh.edu}
\affiliation{Department of Physics \& Astronomy, University of New Hampshire, 9 Library Way, Durham NH 03824, USA}

\author{Zewei Xiong\orcidlink{0000-0002-2385-6771}}
\email{z.xiong@gsi.de}
\affiliation{GSI Helmholtzzentrum {f\"ur} Schwerionenforschung, Planckstra{\ss}e 1, D-64291 Darmstadt, Germany}

\begin{abstract}

Neutrinos in dense astrophysical environments such as core-collapse supernovae (CCSNe) and neutron star mergers (NSMs) can undergo FFCs, which could develop on extremely small scales. A necessary condition for the occurrence of FFCs is the presence of a zero crossing in the electron lepton number (ELN) angular distribution of neutrinos. In this work, we explore machine learning (ML) approaches to detect non-axisymmetric ELN crossings in these environments, based on input features of the $\nu_e$ and $\bar\nu_e$ zeroth and first angular moments. Overall, the ML models demonstrate relatively good generalizability for most of the unseen test datasets generated by various methods that do not assume the same underlying angular distributions as used in the training set. Interestingly, while the model's performance is mediocre for an axisymmetric distribution dataset derived by solving the discretized Boltzmann transport equation under 1D CCSN background, imposing an artificial non-axisymmetry substantially improves the performance. We also find that for the flavor-equilibrated angular distributions, although our ML model trained based solely on ELN inputs performs poorly when the true crossings depend on the post-equilibrated angular distributions of heavy lepton neutrinos and antineutrinos, which become different, it delivers strong performance in detecting ELN crossings when the heavy-lepton neutrino and antineutrino distributions are artificially removed. 
This highlights the need for additional input features to further improve the model. This is a crucial step toward successfully integrating FFCs into large-scale CCSN and NSM simulations.

\end{abstract}

\maketitle

\section{Introduction}
\label{sec:Introduction}
Core-collapse supernovae (CCSNe) and neutron star mergers (NSMs) are among the most extreme astrophysical events in the universe. CCSNe represent the catastrophic death of massive stars, while NSMs arise from the collision of ultra-dense stellar remnants. Both phenomena release enormous amounts of gravitational binding energy in the form of electromagnetic radiations, neutrinos, and gravitational waves, offering unique insights into these most energetic processes in the cosmos. In particular, the intense emission of neutrinos and gravitational waves~\cite{Burrows:2020qrp, Janka:2012wk,Mezzacappa:2026moa,Muller:2026ofz,  Shibata:2019wef,Radice:2020ddv,Foucart:2022bth,Fischer:2023ebq} 
shapes and provides direct information about the complex processes occurring in the cores of these dense objects.

For neutrinos, as they propagate through the dense environments of CCSNe and NSMs, they can experience collective neutrino oscillations~\cite{pantaleone:1992eq, sigl1993general, Pastor:2002we,duan:2006an, duan:2006jv, duan:2010bg, Mirizzi:2015eza,Volpe:2023met,Johns:2025mlm}, driven by coherent forward scattering among themselves. These  interactions produce intricate flavor conversion patterns that can significantly influence the flavor evolution of the neutrino gas and also the dynamics of the explosion. In particular, fast flavor conversions (FFCs) could arise on extremely short scales, characterized by $\sim (G_F n_{\nu})^{-1}$, corresponding to scales as short as a few centimeters within the CCSNe or NSM core~\cite{Sawyer:2005jk,Sawyer:2015dsa,Chakraborty:2016lct, Chakraborty:2019wxe,Capozzi:2020kge, Izaguirre:2016gsx,Capozzi:2017gqd, Wu:2017qpc,Wu:2017drk,Abbar:2017pkh, Capozzi:2018clo, 
Abbar:2018shq, Abbar:2019zoq,George:2020veu,Johns:2019izj, Nagakura:2021hyb,
Abbar:2020qpi,Capozzi:2020syn, DelfanAzari:2019epo,Harada:2021ata,  
Capozzi:2022dtr, Shalgar:2022rjj,    Nagakura:2023wbf, 
Grohs:2023pgq,Akaho:2023brj,Xiong:2024tac,Mukhopadhyay:2024zzl,Nagakura:2025hss,Froustey:2026vhm,Urquilla:2026bff}. 
Here, $G_F$ is the Fermi constant and $n_{\nu}$ is the number density of neutrino. By contrast, conventional slow modes dictated by the neutrino vacuum frequency~\cite{PhysRevD.74.123004, PhysRevD.74.105010, PhysRevLett.97.241101,Padilla-Gay:2025tko,Fiorillo:2025ank,Fiorillo:2025gkw,Froustey:2026vhm} 
and the collision-triggered FCs~\cite{Johns:2021qby,Xiong:2022vsy,Xiong:2022zqz,Liu:2023vtz,Akaho:2023brj,Froustey:2025nbi,Fiorillo:2025zio,Froustey:2026vhm,Urquilla:2026bff}, occur over much larger scales, often spanning a few kilometers. Over the last few years, there has been a growing attempt to implement neutrino FCs in the simulations of CCSNe and NSMs, due to their potential implications on  
neutrino heating, supernova explosion dynamics,  nucleosynthesis outcome, neutrino signals and gravitational wave signals~\cite{Li:2021vqj,Fernandez:2022yyv,Just:2022flt,Ehring:2023lcd, Ehring:2023abs,Qiu:2025ybw,Ehring:2023abs,Akaho:2025giw,Mori:2025cke,Wang:2025nii,Qiu:2025kgy,Wang:2025vbx,Lund:2025jjo,Akaho:2026kff,Gogilashvili:2026epg,Gogilashvili:2026kef}.

A necessary and sufficient condition for occurrence of FFCs is the presence of sign change (zero crossing) in the angular distribution of the difference between neutrinos and antineutrinos, known as the neutrino lepton number ($\nu$LN\footnote{Note that $\nu$LN defined here is the same as the ELN-XLN or (E-X)LN often found in literature.})~\cite{Morinaga:2021vmc,Dasgupta:2021gfs,Fiorillo:2025npi}, which is defined as:

\begin{equation}
\begin{split}
  G(\mathbf{v}) &= \sqrt{2} G_{\mathrm{F}}
  \int_0^\infty \frac{E_\nu^2 \, \mathrm{d} E_\nu}{(2\pi)^3}\,\bigl[(f_{\nu_e}(E_{\nu}, \mathbf{v})- f_{\nu_x}(E_{\nu}, \mathbf{v}))\\
  & \hspace{2.9cm} - (f_{\bar\nu_e}(E_{\nu}, \mathbf{v}) - f_{\bar\nu_x}(E_{\nu}, \mathbf{v}))\bigr],
\end{split}
\label{Eq:G}
\end{equation}
where $\mathbf{v}$ is the neutrino velocity vector, $E_{\nu}$ corresponds to the neutrino's energy, and $f_{\nu}$'s are the neutrino occupation numbers. Here, $\nu_x$ and $\bar\nu_x$ are the heavy flavor neutrinos and antineutrinos. When the $\nu_x$ and $\bar{\nu}_x$ have similar distributions, which could normally be a reasonable approximation, the $\nu$LN reduces to the difference between $\nu_e$ and $\bar{\nu}_e$ distributions, called the electron lepton number (ELN) distribution. 

One of the major challenges in this field is the self-consistent incorporation of FFCs into large-scale hydrodynamical simulations~\cite{Nagakura:2023jfi,Johns:2024dbe,Xiong:2024pue,Johns:2025mlm,Urquilla:2025idk}, as these processes occur on scales much shorter than those currently resolved in such simulations. Moreover, the high computational cost prevents most simulations from providing detailed neutrino angular distributions, making it difficult to directly identify ELN crossings. Instead, neutrino transport in these simulations is commonly approximated by evolving angle-integrated quantities, specifically the lower-order angular distribution moments (zeroth and first order)~\cite{Thorne:1981nvt,Shibata:2011kx,Cardall_2013,Just:2015fda, Murchikova:2017zsy,  Tejedor:2025xxx} defined as,

\begin{equation}
\begin{split}
\mathrm{I_{0}} &= \int_{-1}^1 \mathrm{d}\cos\theta_\nu 
\,\int_0^{2\pi} \mathrm{d}\phi_{\nu}
\,\int_0^\infty \frac{E_\nu^2 \mathrm{d} E_\nu}{(2\pi)^3}
\, f_{\nu}(E_{\nu}, \mathbf{v}), \\
\mathbf{I_{1}} &= \int_{-1}^1 \mathrm{d}\cos\theta_\nu 
\,\int_0^{2\pi} \mathrm{d}\phi_{\nu}
\,\int_0^\infty \frac{E_\nu^2 \mathrm{d} E_\nu\, \mathbf{v}}{(2\pi)^3}
\, f_{\nu}(E_{\nu}, \mathbf{v}),
\end{split}
\label{Eq:In}
\end{equation}
where $\mathbf{v} =  (\sin{\theta_{\nu}} \cos{\phi_{\nu}},\  \sin{\theta_{\nu}} \sin{\phi_{\nu}},\ \cos{\theta}_{\nu})$, with $\phi_\nu$ and $\theta_\nu$ being the azimuthal and zenith angles of the neutrino momentum with respect to z-axis.

Although the use of a limited set of neutrino angular moments, rather than full angular distributions, entails a loss of information, several methods have been proposed to infer the presence of angular crossings from the available $\mathrm{I_0}$ and $\mathbf{I_1}$ in simulations. Early works in this direction predominantly employed analytical and semi-analytical approaches~\cite{Dasgupta:2018ulw,Johns:2021taz,Johns:2019izj,Abbar:2020fcl,Nagakura:2021hyb,Johns:2021taz, Richers:2022dqa}. 
In particular,~\cite{Richers:2022dqa} presented a study comparing the different methods to identify ELN crossings using the Monte Carlo transport data of a binary NSM. Similar comparison has also been recently carried out in~\cite{Cornelius:2025tyt} in the context of CCSN models.

It turns out that machine learning (ML) approaches~\cite{Abbar:2023kta,Abbar:2023zkm,Abbar:2024chh} can also act as a potentially powerful tool for identifying FFCs in CCSN and NSM simulations. These methods learn patterns from large datasets and can generalize to previously unseen inputs without requiring explicit programming. Although they require substantial training data, they offer significant advantages in post-training speed and computational efficiency. In particular, once trained, such models may be efficiently  deployed on the fly within CCSN and NSM simulations, in contrast to more computationally intensive traditional approaches. Pioneer studies in this direction~\cite{Abbar:2023kta,Abbar:2023zkm} focused primarily on detecting axisymmetric ELN crossings, i.e., those depending only on the zenith angle. This limitation largely arises from the absence of a clear analytical understanding of the functional form of the neutrino angular distribution in the azimuthal direction. However, $\phi$-dependent (non-axisymmetric) crossings are expected to be prevalent in such environments. Motivated by this, recent work~\cite{Abbar:2024chh} has explored the detection of non-axisymmetric ELN crossings using ML, with models trained and tested on simulation data from both rotating and non-rotating CCSN. 

In this work, we extend these efforts by developing a training framework based on systematically generated synthetic data that spans a broad and general parameter space beyond the limitations of simulation-driven datasets. We construct non-axisymmetric configurations by introducing a relative frame rotation between the neutrino and antineutrino angular distributions, each described by an underlying parametric form. The resulting ML models demonstrate strong performance in identifying ELN crossings across datasets drawn from varied sources. Furthermore, our results emphasize the importance of maintaining consistency between the input feature spaces of the training and test datasets for reliable generalization of the ML model.

In Section~\ref{sec: Framework}, we describe our framework, detailing the input features, training data, and ML algorithms employed. Section~\ref{sec: test datasets} presents the performance of the models on a variety of test datasets obtained from different sources. Finally, we summarize our results and discuss their implications in Section~\ref{sec:Discussion}.
\section{Framework}
\label{sec: Framework}
In this section, we present the setup used to generate the training dataset which are used by our ML models to detect the ELN crossings. Following the approach adopted in~\cite{Abbar:2023kta}, we employ the maximum entropy angular distribution as the underlying form for both neutrinos and antineutrinos. Note that we obtain similar results when a gaussian distribution is used instead. The choice of a parametric form provides a generic and minimally constrained representation of the angular distributions, allowing us to systematically explore a wide range of possible ELN configurations. The general axisymmetric form of the distribution is given by:
\begin{equation}
g_{\nu}(\mathrm{v}_z) = \exp[-\eta + a\,\mathrm{v}_z],
\label{Eq:fnu}
\end{equation}
where $g_{\nu}(\mathrm{v}_z)$ is defined as:
\begin{equation}
g_{\nu}(\mathrm{v}_z) = \int_0^{2\pi}\,\mathrm{d}\phi_{\nu}\,\int_0^\infty\,\frac{E_{\nu}^2\,\mathrm{d}E_{\nu}}{(2\pi)^3}f_{\nu}(E_{\nu}, \mathbf{v}).
\label{Eq:gnu}
\end{equation}
Here $\mathrm{v}_z = \cos{\theta_{\nu}}$, and $\eta$ and $a$ determine the overall neutrino number density and the shape of the distribution, respectively. Note that here we consider only the angular distributions of $\nu_e$ and $\bar\nu_e$. The heavy flavors, namely $\nu_x$ and $\bar\nu_x$, are assumed to have similar distributions and thus are not relevant for the detection of crossings. 

In this work, our aim is to investigate the scenarios where the angular distributions depend on both the zenith ($\theta_{\nu}$) and azimuthal ($\phi_{\nu}$) angles of the neutrino momentum, allowing us to capture non-axisymmetric ELN crossings. To achieve this, we construct configurations in which the propagation directions of neutrinos and antineutrinos are rotated relative to each other, thereby introducing non-axisymmetric ELN distributions. Without loss of generality, we assume that the first moment of $\nu_e$ only has a nonvanishing component in $z$ direction, and rotate the antineutrino angular distribution about the y-axis by an arbitrary angle $\theta_r$. While $g_{\nu_e}$ remains unchanged, the $\bar\nu_e$ angular distributions  take the form: 
\begin{equation}
g_{\bar{\nu}_e}(\mathrm{v}_z') = \exp\!\left[-\eta_{\bar{\nu}_e} + a_{\bar{\nu}_e} \mathrm{v}_z'\right],
\label{Eq:fnu'}
\end{equation}
where $\mathrm{v}_z' = \sin{\theta_r} \mathrm{v}_x + \cos{\theta_r} \mathrm{v}_z$ with $\mathrm{v}_x = \sin{\theta_{\nu}} \cos{\phi_{\nu}}$ and $\mathrm{v}_z = \cos{\theta}_{\nu}$.
\subsection{Input Features}
\label{sec:Input Features}
To determine the presence of ELN crossings, we utilize the zeroth and first moments of the neutrino and antineutrino distributions, $\mathrm{I_0}$ and $\mathbf{I_1}$ (Equation~(\ref{Eq:In})). Rather than treating each moment independently, we consider the specific ratios which encapsulate the relevant information about the crossings. This is based on the fact that the existence of crossings does not depend on the overall normalization factor. The four input features used for training our model are defined as:
\begin{equation}
\begin{aligned}
\alpha &= \frac{\mathrm{I}^{\bar\nu_e}_0}{\mathrm{I}^{\nu_e}_0}, \quad 
F_{\nu_{e}z} &= \frac{\mathrm{I}^{\nu_e}_{1z}}{\mathrm{I}^{\nu_e}_0}, \\
F_{\bar\nu_e x} &= \frac{\mathrm{I}^{\bar\nu_e}_{1 x}}{\mathrm{I}^{\bar\nu_e}_0}, \quad 
F_{\bar\nu_e z} &= \frac{\mathrm{I}^{\bar\nu_e}_{1 z}}{\mathrm{I}^{\bar\nu_e}_0},
\end{aligned}
\label{Eq:alpha}
\end{equation}
where $\alpha$, the ratio of the number densities of $\bar\nu_e$ and $\nu_e$, quantifies the relative abundances of the two species. The quantities $F_{\nu_e/\bar\nu_e}$ are the corresponding flux factors, with the subscripts x and z indicating their directions. Note that $F_{\bar\nu_e y}=0$ as the rotation is done about the y-axis.
\subsection{Training data}
\label{sec:Training data}
Our fiducial model is trained by systematically generating the above mentioned input features and the corresponding crossing labels in order to create a generic dataset. We randomly generated 100 values of $\theta_r$ linearly in the range (0, $\pi$/2). For each $\theta_r$, we sampled 100 values of $\alpha$ logarithmically in the range (0.001, 10). Subsequently, for each pair of ($\theta_r$, $\alpha$), 10 random values were generated for each of the flux factors, $F_{\nu_ez}$, $F_{\bar\nu_ex}$ and $F_{\bar\nu_ez}$. This procedure yields a total of $10^5$ data points in the input dataset. Using these sampled features, we reconstruct the angular distributions according to Equations~\eqref{Eq:fnu} and \eqref{Eq:fnu'} and determine the presence or absence of ELN crossings, which serve as the ground-truth labels for training. We further partition the dataset such that 70$\%$ is used for training, while the remaining 30$\%$ is equally split between validation and test sets. The training set is used to fit the model parameters, whereas the validation set is employed for hyperparameter tuning based on the model’s performance. After selecting the optimal hyperparameters, the model is retrained on the combined training and validation sets, and its final performance is evaluated on the test set.

\begin{figure}[t!] 
    \centering
    \includegraphics[width=\columnwidth]{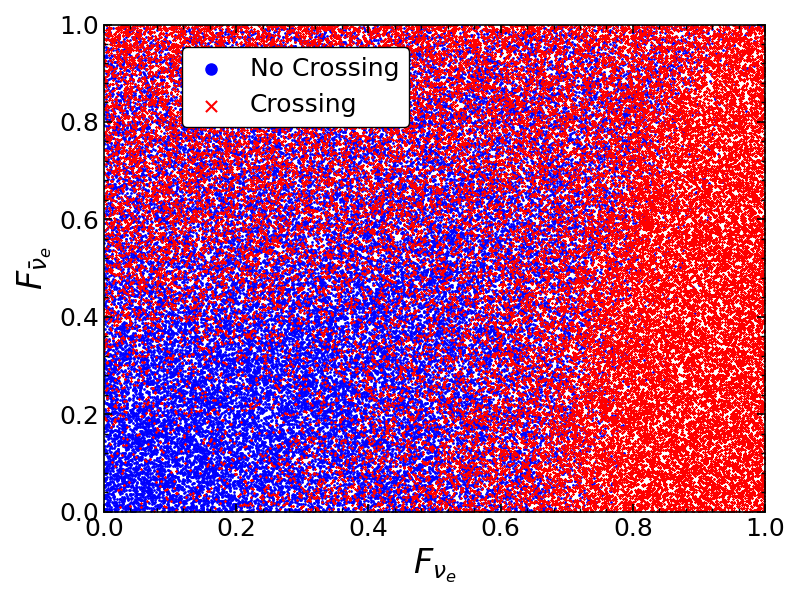}
    \caption{crossings/no crossings in the $F_{\nu_e}- F_{\bar\nu_e}$ space for the entire training dataset. Note that the points are plotted for all the values of $\alpha$ in the training dataset.}
    \label{fig:param_space maxent}
\end{figure}

In Figure~\ref{fig:param_space maxent}, we illustrate the regions with and without crossings in the two-dimensional  space spanned by the magnitude of the neutrino and antineutrino flux factors, $F_{\nu_e}$ and $F_{\bar \nu_e}$, for the full training dataset. Note that it includes the flux factors corresponding to all values of $\alpha$ in the dataset. We find that our training dataset is quite balanced, i.e., $49.7\%$ of the samples exhibit crossings, while the remaining $50.3\%$ do not. 

We formulate the detection of ELN crossings as a binary classification problem, where class 0 corresponds to the absence of a crossing and class 1 indicates the presence of a crossing. To evaluate the performance of the ML models, we employ standard classification metrics, namely accuracy, precision, recall, and the $F_1$ score, which are defined as follows:
\begin{equation}
\begin{split}
&\mathrm{accuracy} = \frac{\text{T}_\text{p} + \text{T}_\text{n}}{\text{T}_\text{p} + \text{T}_\text{n} + \text{F}_\text{p} + \text{F}_\text{n}}, \\
&\mathrm{precision} = \frac{\text{T}_\text{p}}{\text{T}_\text{p}+\text{F}_\text{p}}, \\
&\mathrm{recall} = \frac{\text{T}_\text{p}}{\text{T}_\text{p}+\text{F}_\text{n}}, \\
&F_1 = 2\times \frac{\mathrm{precision} \times \mathrm{recall} }{\mathrm{precision} + \mathrm{recall}},
\end{split}
\end{equation}
where T and F represent the true and false classifications and the subscripts p and n denote the positive and negative classifications, respectively. The accuracy tells us how many predictions are correct out of the total number of predictions. This works best when both classes are equally important, like in our case, where we want our model to be sensitive to both the presence and absence of crossings. Precision provides the proportion of the positive predictions that are actually correct while recall identifies the fraction of actual positives which are correctly predicted. $F_1$ score is the harmonic mean of precision and recall which balances both the metrics.  
\subsection{ML algorithms}
\label{sec:ML Algorithms}
In this section, we discuss the four ML algorithms employed in our analysis, namely Logistic Regression~(LR), k-Nearest Neighbors~(KNN), Support Vector Machine~(SVM) and Decision Tree~(DT). A brief overview of each of these methods is provided below:

\begin{enumerate}
\item Logistic Regression : It is one of the simple and most widely used algorithms for the purpose of binary classifications. In this method, a sigmoid function is  applied to a linear combination of the input features as follows:
\begin{equation}
\sigma (z)={\frac{1}{1+e^{-z}}},
\label{Eq:sigma}
\end{equation}
where $z = \vec{w} \cdot \vec{x}$; $\vec{x}$ and $\vec{w}$ are the vectors representing the input features and their corresponding weights, respectively. This function produces a value between 0 and 1, which represents the predicted probability of the positive class. Based on a threshold value (taken to be 0.5 here), the model classifies the input into one of the two classes. In other words, if $\sigma (z)>0.5$, the input is classified into class 1 (crossing) and class 0 (no crossing) if $\sigma (z)<0.5$.  
However, in order to effectively apply this classifier to our problem, we generate additional features by performing non-linear transformations on the original inputs, $\vec{x}$. This preprocessing is carried out by using the tools provided by the Python scikit-learn library. 

The top panel of Figure~\ref{fig:hyperparameters} shows the variation in precision and accuracy with the degree of polynomial ($n$) used in the LR algorithm for training and validation datasets. One can see that the highest accuracy and precision are achieved for a polynomial of degree $n=4$ in both cases, after which performance decreases. This behavior can be understood in terms of the bias–variance tradeoff. For $n<4$, the number of polynomial features is insufficient to capture the underlying complexity of the model, leading to underfitting. At $n=3$ or 4, the model complexity is well matched to the data, resulting in optimal performance. Beyond
$n>4$, the model begins to overfit, capturing noise in the training data rather than the true underlying patterns, which causes a decline in validation accuracy and precision. So, we choose $n=4$ for our analysis.

\item K-Nearest Neighbors : In this algorithm, data points are classified based on the classes of the $k$ nearest neighbors in the training set. This is a non-parametric, instance-based method, i.e., it does not learn an explicit model during training. Instead, it makes predictions by comparing the new data point to the entire training dataset identifying the most similar points. While this approach is simple and effective, it can become computationally inefficient for large datasets, as the algorithm must scan the entire dataset each time a prediction is made.  

The middle panel of Figure~\ref{fig:hyperparameters} shows the variation of accuracy and precision with the number of nearest neighbors for the training and validation datasets. Here, one can see that when the model is tested on the training set, 100$\%$ accuracy and precision is achieved but they are less for the validation set. The highest accuracy and precision is achieved for $k = 7$ and further when the model is tested on the validation set. So, we choose $k = 7$ for our further analysis. 

\item Support Vector Machine : The main idea behind this algorithm is to find the best decision boundary (known as a hyperplane) separating the different classes in the dataset. It does so in a way such that the distance between the data points in different classes and the hyperplane is maximised.  In a simple 2D case, this hyperplane is a straight line which separates the data points of different classes. However, when the data is not separable linearly, a technique known as kernel trick is used to map it to a higher dimensional space where linear separation is possible. We use the radial basis function (RBF) kernel in our analysis which is defined as:
\begin{equation}
\mathcal{K}(\vec{x},\vec{x}') = \exp(-\gamma ||\vec{x}-\vec{x}'||^2),
\label{Eq:RBF}
\end{equation}
where $\vec{x}$ and $\vec{x}'$ are the test and training data points, respectively. The RBF kernel measures the similarity between two points and $\gamma$ controls how quickly the similarity drops off with distance. 

\begin{figure}[tbh!]
    \centering

    \begin{subfigure}[b]{\columnwidth}
        \includegraphics[height=5cm,keepaspectratio]{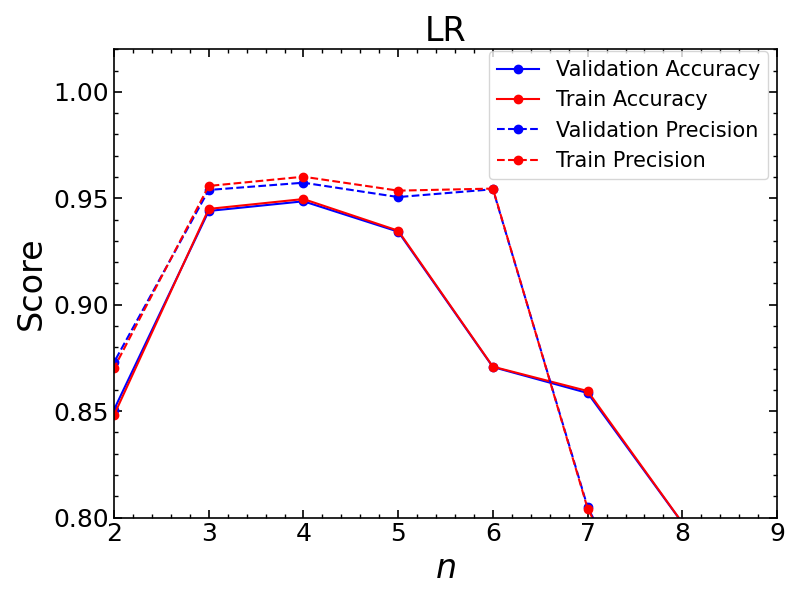}
    \end{subfigure}

    \vspace{0.3cm}

    \begin{subfigure}[b]{\columnwidth}
        \includegraphics[height=5cm,keepaspectratio]{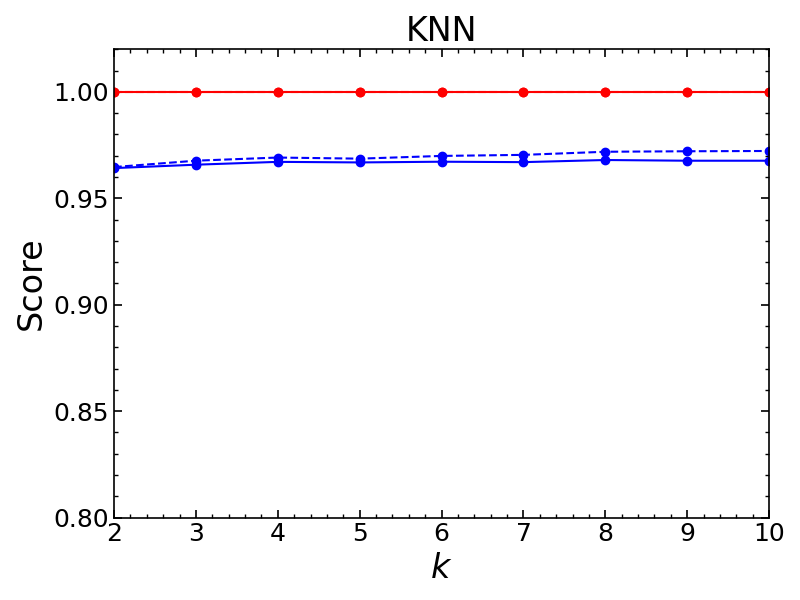}
    \end{subfigure}

    \vspace{0.3cm}

    \begin{subfigure}[b]{\columnwidth}
        \includegraphics[height=5cm,keepaspectratio]{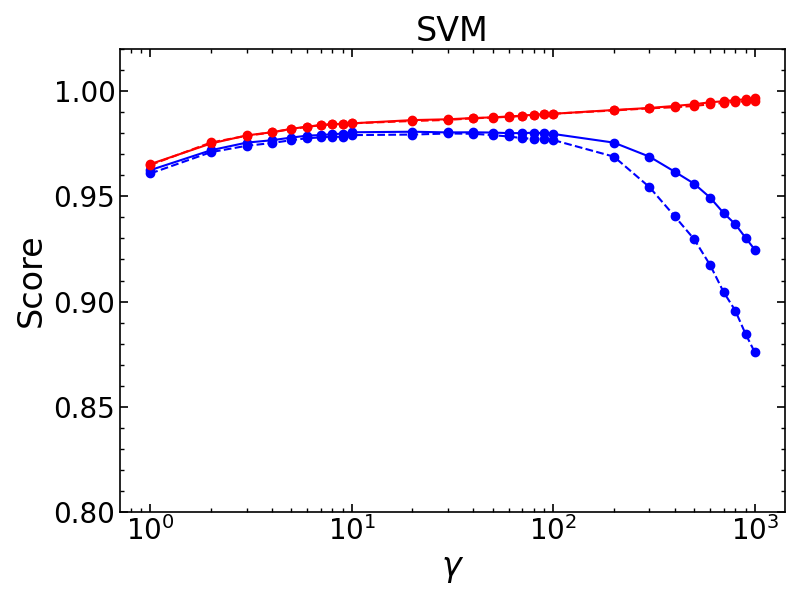}
    \end{subfigure}
    \caption{Precision and accuracy scores of the LR, KNN and SVM algorithms evaluated on the training dataset (used as a test set) and on the validation dataset. The top panel represents the variation of metrics with $n$ used for preprocessing in LR. The middle panel refers to the variation of metrics with the $k$ used in KNN. The bottom panel is for the variation of metrics with parameter $\gamma$ used in SVM.}
    \label{fig:hyperparameters}
\end{figure}

The bottom panel of Figure~\ref{fig:hyperparameters} shows the variation of the accuracy and precision scores with the parameter $\gamma$ for training and validation datasets. As can be seen from the figure, both the accuracy and precision for the validation set reach maximum at $\gamma = 10^2
$ which we choose as the fiducial value for our further analysis.

\item Decision Tree Classifier : It is a widely used supervised learning algorithm for binary classification tasks. It builds a tree-like model of decisions by recursively splitting the dataset based on input feature values. At each node, the algorithm selects the feature that best separates the two classes, continuing the process until the data is sufficiently classified. Its intuitive structure and interpretability make it a useful tool for understanding how input features contribute to the final prediction. However, decision trees are prone to overfitting, especially when the tree becomes too deep or when the dataset contains noise. Note that here we implement this method using the default hyperparameter settings provided by sci-kit learn.
\end{enumerate}
\section{Test datasets}
\label{sec: test datasets}
In this section, we evaluate the performance of the trained models on a variety of datasets drawn from different sources, with the aim of examining their capability to generalize beyond the training distributions. In particular, we investigate how well the models can identify the presence or absence of ELN crossings across diverse physical scenarios.
\subsection{Angular distributions from same method as training dataset}
\label{sec:Dataset from same method as training set}
We begin by testing our trained ML models on the test dataset generated in Section~\ref{sec:Training data}. Our test set contains 15000 sample points. The performance metrics for the different ML algorithms are shown in Table~\ref{tab:tab_test}. We see that overall our model performs well across all methods, with accuracies exceeding 95$\%$ in each case. These high values of accuracies are as expected because our test dataset has been generated by the same method as the training set. 
\begin{table}[tbh!]
\centering
\caption{The performance metrics of the different ML algorithms trained and tested on the datasets mentioned in Section~\ref{sec:Training data}.}
\begin{tabular}[t]{|lcc|c|}
\hline
& \textcolor{black}{ \textbf{{Logistic Regression}}} (95\%) \\
\hline
& precision & recall & $F_1$-score \\
\hline
no  crossing & 95\% & 95\% & 95\% \\
 crossing& 95\% & 95\% & 95\% \\
\hline
& \textcolor{black}{ \textbf{{KNN}}} (97\%) \\
\hline
& precision & recall & $F_1$-score \\
\hline
no  crossing & 97\% & 97\% & 97\% \\
 crossing& 97\% & 97\% & 97\% \\
\hline
& \textcolor{black}{ \textbf{{SVM}}} (98\%) \\
\hline
& precision & recall & $F_1$-score \\
\hline
no  crossing & 98\% & 98\% & 98\% \\
 crossing& 98\% & 98\% & 98\% \\
\hline
& \textcolor{black}{ \textbf{{Decision Tree}}} (97\%) \\
\hline
& precision & recall & $F_1$-score \\
\hline
no  crossing & 97\% & 97\% & 97\% \\
 crossing& 97\% & 97\% & 97\% \\
\hline
\end{tabular}
\label{tab:tab_test}
\end{table}

\subsection{Ray-traced angular distributions above the
\texorpdfstring{$\nu_e$ and $\bar \nu_e$}{nue and nubar_e} 
emission surfaces in a binary NSM remnant}
\label{sec:NSM ray-tracing}

We now test our trained ML models on a dataset in which the angular distributions of $\nu_e$ and $\bar\nu_e$ are obtained from  the ray-tracing method as described in Ref.~\cite{PhysRevD.96.123015}. At each point on the emission surface, the neutrino emission is characterized by the angle $\theta_N$, defined as the angle between the emission direction and the local surface normal. The angular dependence of the neutrino flux is then assumed to follow:
\begin{equation}
\Phi_{\nu} (\theta_N) = \frac{n_{\nu} (1 + \cos{\theta_N})}{4\pi},
\end{equation}
where $n_{\nu}$ is the number density of neutrinos (or antineutrinos) at the emission point. To obtain the angular distributions at locations above the emission surface, we perform ray tracing of neutrino intensities from all the points on the emission surface to the target location.

Here, we consider the neutrino angular distributions obtained by the ray-tracing method in a NSM remnant examined in Ref.~\cite{George:2020veu} for 1000 randomly chosen locations above the emission surfaces in the $x$-$z$ plane with  $x \in (0-100)$ and $z\in (25-100)$~km. In addition, the neutrino number densities (at the emission surfaces) were obtained from the fully 3D NSM simulations~\cite{10.1093/mnras/stz613} of the mergers of two non-rotating NSs with mass 1.35 $M_\odot$ each, with the EoS of DD2~\cite{PhysRevC.81.015803}, where we specially used the azimuthally averaged simulation data at 2.5 ms after the coalescence.
Our input features span the ranges $\alpha \in [1.66, 2.05]$, $F_{\nu_e}\in [0.66, 0.98]$, and $F_{\bar \nu_e}\in[0.72, 0.99]$, where $F_{\nu}$ is the magnitude of the (anti)neutrino flux factor.

The performance metrics for this case are shown in Table~\ref{tab:NSM maxent}. Note that by construction, all the cases in our test set contain crossings, i.e., it is a biased set. As a result, we can see that the precision is always $100\%$ by definition, and the recall is equal to the accuracy. Note also that while the performance in identifying crossings is nearly perfect independent of the ML algorithms, due to the biased sample, our results here cannot test whether our ML model can well classify non-crossing distributions generated by the same method.
\begin{table}[tbh!]
\centering
\caption{The performance metrics of the different ML algorithms trained on our training dataset (Section~ \ref{sec:Training data}) and tested on the dataset generated by ray-tracing method in a NSM remnant. Note that in the test set all the points present crossings.}
\begin{tabular}[t]{|lcc|c|}
\hline
& \textcolor{black}{ \textbf{{Logistic Regression}} (100\%)}    \\
\hline
& precision & recall & $F_1$-score \\
\hline
 crossing&100\%&100\% & 100\% \\
\hline
&\textcolor{black}{  \textbf{{KNN (k=7)}}} (100\%) \\
\hline
&precision&recall & $F_1$-score \\
\hline
 crossing&100\%&100\%&100\%\\
\hline
&\textcolor{black}{\textbf{{SVM}}} (100\%)\\
\hline
&precision&recall & $F_1$-score\\
\hline
 crossing&100\%&100\%&100\%\\
\hline
&\textcolor{black}{\textbf{{Decision tree}}} (93\%)\\
\hline
&precision&recall & $F_1$-score \\
\hline
 crossing&100\% & 93\% & 96\%\\
\hline
\end{tabular}
\label{tab:NSM maxent}
\end{table}

\subsection{Angular distributions from 1D CCSN simulations}
\label{sec:Zewei's simulation}

In this section, we consider a set of $\nu_e$ and $\bar \nu_e$ angular distributions 
used in Ref.~\cite{Xiong:2024tac} without considering FFCs, which were obtained by solving the classical neutrino transport equations over several static and 1D CCSN background profiles, adjusted from a CCSN snapshot from an 1D CCSN simulation with a 25~$M_\odot$ progenitor star computed using the \texttt{AGILE-BOLTZTRAN} code~\cite{Liebendoerfer:2002xn,Fischer:2020vie}. 
The range of the input features are $\alpha \in [0.46, 1.28]$, $F_{\nu_e} \in [0.007, 0.86]$, and $F_{\bar \nu_e} \in [0.009, 0.84]$.

We then test our ML model on two datasets obtained from the original one. In the first one, the original axisymmetric distributions is used and in the second one, non-axisymmetric angular distributions are generated by rotating 
the $\bar \nu_e$ angular distribution about the y-axis by an angle chosen as $\theta_r = \pi/6$. Here, we have 392 such sample points for each of the datasets.

We first test our trained model on the axisymmetric angular distributions, where we have 190 samples for crossings and 202 for no crossings. The  performance metrics are presented in Table~\ref{tab:Zewei's sim_axisymm}. We notice that the overall performance of all the ML algorithms is not that good. This is expected as these angular distributions are obtained from the realistic simulations which can be quite different from those with the parametric form  used to train the model. One interesting thing to note here is that the algorithms LR, KNN and SVM do a better job in identifying all the actual crossings, which is evident from the recall but they fail to detect efficiently the cases without crossings.

Now we focus on how our model performs when tested on the dataset containing the non-axisymmetric features as well, where there are 303 crossing samples and 89 no crossing ones. Table~\ref{tab:Zewei's sim rotated} shows the performance metrics for this dataset. We notice that the performance of our ML model is interestingly much better than that of the axisymmetric one. Overall, the algorithms do a great job in detecting the crossings compared to the no crossings, with even higher recalls than in the axisymmetric scenario. In particular, the recalls are $100\%$ for LR and SVM.

We find that in the $F_{\nu_e}-F_{\bar\nu_e}$ parameter space for the test dataset, the two classes are better separated in the non-axisymmetric case compared to the axisymmetric case. In addition, the fraction of distributions exhibiting crossings also increases when we move from the axisymmetric to the non-axisymmetric case. Correspondingly, we observe a reduction in the misclassification rates for both crossing and no-crossing cases. For example, in the axisymmetric case, LR misclassifies approximately $70\%$ of the no crossing distributions. In contrast, for the non-axisymmetric case, these misclassification rates decrease significantly to around $57\%$. Both these factors are perhaps the reason why we have obtained better ML performance in the non-axisymmetric case.  

\begin{table}[tbh!]
\centering
\caption{The performance metrics of our models (trained in Section~\ref{sec:Training data}) tested on the dataset generated by using the axisymmetric angular distributions provided by 1D SN simulations (Section~\ref{sec:Zewei's simulation}). Note that the overall accuracies are low, with that of DT being the worst.}
\begin{tabular}[t]{|lcc|c|}
\hline
& \textcolor{black}{ \textbf{{Logistic Regression}} (64\%)}   \\
\hline
& precision & recall & $F_1$-score \\
\hline
no  crossing & 100\% & 30\% & 46\% \\
crossing& 57\% & 100\% & 73\% \\
\hline
&\textcolor{black}{  \textbf{{KNN (k=7)}}(67\%)} \\
\hline
&precision&recall & $F_1$-score \\
\hline
no  crossing & 93\% & 40\% & 56\% \\
crossing& 60\% & 97\% & 74\% \\
\hline
&\textcolor{black}{   \textbf{{SVM}} (68\%)}\\
\hline
&precision&recall & $F_1$-score\\
\hline
no  crossing & 95\% & 40\% & 56\% \\
crossing& 60\% & 98\% & 75\% \\
\hline
&\textcolor{black}{   \textbf{{Decision tree}} (52\%)}\\
\hline
&precision&recall & $F_1$-score \\
\hline
no  crossing & 54\% & 42\% & 47\% \\
crossing& 50\% & 62\% & 55\% \\
\hline
\end{tabular}
\label{tab:Zewei's sim_axisymm}
\end{table}%

\begin{table}[tbh!]
\centering
\caption{Same as Table~\ref{tab:Zewei's sim_axisymm}, but for the artificially generated nonaxisymmetric distributions. Note that our ML model performs much better on this test dataset compared to that of Table~\ref{tab:Zewei's sim_axisymm}. In particular, the recall scores show the model is able to classify the crossings almost perfectly.}
\begin{tabular}[t]{|lcc|c|}
\hline
& \textcolor{black}{ \textbf{{Logistic Regression}} (87\%)}    \\
\hline
& precision & recall & $F_1$-score \\
\hline
no  crossing & 100\% & 43\% & 60\% \\
crossing& 86\% & 100\% & 92\% \\
\hline
&\textcolor{black}{  \textbf{{KNN (k=7)}} (92\%)} \\
\hline
&precision&recall & $F_1$-score \\
\hline
no  crossing & 91\% & 70\% & 79\% \\
crossing& 92\% & 98\% & 95\% \\
\hline
&\textcolor{black}{  \textbf{{SVM}} (92\%)} \\
\hline
no  crossing & 98\% & 65\% & 78\% \\
crossing& 91\% & 100\% & 95\% \\
\hline
&\textcolor{black}{   \textbf{{Decision tree}} (93\%)}\\
\hline
&precision&recall & $F_1$-score \\
\hline
no  crossing & 92\% & 74\% & 82\% \\
crossing& 93\% & 98\% & 95\% \\
\hline
\end{tabular}
\label{tab:Zewei's sim rotated}
\end{table}%

\subsection{Angular distributions from Monte Carlo data of a binary NSM}
\label{sec:MC simulation}

In this section, we test our trained model on a dataset extracted from a time snapshot at 4~ms post merger in a binary NSM simulation that adopts the Monte Carlo (MC) neutrino transport~\cite{foucart2024robustnessneutronstarmerger}.
At a given time snapshot, each MC packet represents a  particular number of neutrinos that have the same four momentum and are at the same spatial coordinates. 
Since different packets have different momenta and locations, we perform volume averaging to approximately reconstruct the corresponding neutrino angular distribution at different locations~\cite{Mukhopadhyay:2024zzl} as follows. 
At the location of our interest, we consider a cubical box centered at that point, whose size is chosen such that at least 10000 MC packets separately for $\nu_e$ and $\bar\nu_e$ are enclosed. 
Then, we discretize the angular domain into 12 bins each in $\theta_\nu$ and $\phi_\nu$, respectively\footnote{We have checked that taking a different minimal number of MC packets or taking different angular bins do not substantially affect the obtained angular distributions.}.
For each angular grid, we sum over all MC packets whose momentum directions fall within that grid. 
This procedure allows to generate relatively smooth (although still discretized) neutrino angular distributions for each chosen location. 
These angular distributions are then used to compute the corresponding moments and to determine whether ELN crossings exist or not.
We have considered 2225 discrete spatial locations with radial distances ranging from approximately 20 km to 118 km from the center of the merger remnant, providing samples from both relatively near and relatively far regions of the remnant. The range of input features are $\alpha \in [2.27, 28.3]$, $F_{\nu_e}\in [0.08, 0.71]$, and $F_{\bar\nu_e}\in[0.09,0.76]$. 
We find that all the cases are without crossing which makes it a biased dataset. This could be related to the overall significant dominance of the $\bar\nu_e$ number density over that of $\nu_e$. Furthermore, owing to the limited number of MC packets, volume averaging is used to reconstruct the angular distributions with sufficient statistical accuracy. This averaging mixes packets from neighboring regions with different angular distributions, thereby smoothing local ELN features and potentially suppressing ELN crossings.

The performance metrics of our ML model for this dataset are presented in Table~\ref{tab:NSM MC maximum entropy}. We see that all the algorithms can make predictions with $100\%$ precision which is expected in this case as it has labels corresponding to only one class. In this case, the algorithm SVM behaves in a different manner compared to others. It can correctly predict a no crossing  only $11\%$ of the times, and misidentifies it as a crossing majority of the times.
\begin{table}[tbh!]
\centering
\caption{The performance metrics of our models (trained in Section~\ref{sec:Training data}) tested on the dataset generated by using the Monte carlo data of a NSM simulation.}
\begin{tabular}[t]{|lcc|c|}
\hline
& \textcolor{black}{ \textbf{{Logistic Regression}} (100\%)}    \\
\hline
& precision & recall & $F_1$-score \\
\hline
no  crossing &100\%  & 100\% & 100\% \\
\hline
&\textcolor{black}{  \textbf{{KNN (k=7)}} (98\%)} \\
\hline
&precision&recall & $F_1$-score \\
\hline
no  crossing &100\%  & 98\% & 99\% \\
\hline
&\textcolor{black}{   \textbf{{SVM}} (11\%)}\\
\hline
&precision&recall & $F_1$-score\\
\hline
no  crossing &100\%  & 11\% & 20\% \\
\hline
&\textcolor{black}{   \textbf{{Decision tree}} (99\%)}\\
\hline
&precision&recall & $F_1$-score \\
\hline
no  crossing &100\%  & 99\% & 99\% \\
\hline
\end{tabular}
\label{tab:NSM MC maximum entropy}
\end{table}
\subsection{Angular distributions from applying flavor equilibration to the initial input dataset}
\label{sec: FE scheme}

Several studies~\cite{Bhattacharyya:2020dhu,Bhattacharyya:2020jpj,Wu:2021uvt,Richers:2021nbx,Zaizen:2021wwl,Richers:2021xtf,Bhattacharyya:2022eed,Grohs:2022fyq,Abbar:2021lmm,Richers:2022bkd,Zaizen:2022cik,Xiong:2023vcm,DelfanAzari:2024xgs,Richers:2024zit,George:2024zxz,Liu:2025tnf,Liu:2025muc,Goimil-Garcia:2025ozm,Fiorillo:2026byh} have explored the final outcome of FFCs using local box simulations with periodic boundary conditions. These works demonstrate that the system evolves toward quasi-stationary states with survival probabilities governed by the neutrino lepton number conservation. Analytical prescriptions have been formulated to describe these probabilities, and they show good agreement with numerical simulation results~
\cite{Bhattacharyya:2020jpj,Zaizen:2022cik,Xiong:2023vcm,Xiong:2024pue,Richers:2024zit,George:2024zxz}. 
In order to assess the performance of our models on 
non-Gaussian distributions
that correspond to these post-FFC equilibrated states
we take the test set described in Section~\ref{sec: Framework}, and apply the ``box-like" prescription described in ~\cite{Zaizen:2022cik,Xiong:2023vcm} to compute the equilibrated angular distributions as follows. 
The area of the positive and negative regions in the original ELN angular distribution are given by the integrals $I_+$  and $I_-$, defined as,
\begin{equation}
\begin{aligned}
    I_+ &= \left|\int \mathrm{d}\Gamma \ G(\mathrm{v}_z,\phi_\nu) \ \Theta[G(\mathrm{v}_z, \phi_\nu)] \right|, \\
    I_- &= \left|\int \mathrm{d}\Gamma \ G(\mathrm{v}_z,\phi_\nu) \ \Theta[-G(\mathrm{v}_z, \phi_\nu)] \right|,
    \end{aligned}
\end{equation}
where $G(\mathrm{v}_z, \phi_\nu)$ is the ELN angular distribution, $\Theta$ is the Heaviside theta function, and $\int \mathrm{d}\Gamma = \int_{-1}^{1} \mathrm{dv}_z \ \int_0^{2\pi} \mathrm{d}\phi_\nu $. The depth of the crossing is represented by a quantity $I_{\rm ratio} = I_</I_>$, where $I_{<} =\,$min$(I_{-},I_{+})$, and  $I_{>} =\,$max$(I_{-},I_{+})$. Assuming that the small side of the ELN angular distribution undergoes complete flavor equilibration, the asymptotic survival probability of $\nu_e$ is given by:   
\begin{equation}
    P_{ee}(\Gamma) = \begin{cases}
\frac{1}{2}, & \text{for}\,\Gamma^< \,, \\
1-\frac{I_<}{2\,I_>}, & \text{for}\,\Gamma^> \,,
\end{cases} 
\label{eq : FE}
\end{equation}
where $\Gamma^<$ (small side) and $\Gamma^>$ (large side) are the ($\mathrm{v}_z, \phi_\nu$) regions corresponding to $I_<$, and $I_>$, respectively.

The performance metrics of the different ML algorithms on the
 dataset generated
after applying flavor equilibration to the neutrino
angular distributions
 are summarized in Table~\ref{tab:FE maxent}, where all models perform poorly, as might be expected. This can be traced to the limited input features, which include only $\nu_e$ and $\bar\nu_e$. However, FFCs can lead to very different fluxes of $\nu_x$ and $\bar{\nu}_x$, making the inclusion of heavy-lepton neutrino information essential. In particular, the true labels for crossings/no crossings are determined using the 
 $\nu$LN 
 defined by Equation~\eqref{Eq:G} which depends on all neutrino species, whereas the model inputs are restricted to $\nu_e$ and $\bar\nu_e$. This inconsistency leads to the misclassification of the true no crossing cases as crossings.       

In order to have consistency in the provided features to the ML models, we now construct a dataset in which the input features remain identical to those described previously, but the true labels are determined solely by the ELN, i.e., without accounting for the heavy-flavor neutrino fluxes. We find that all crossings present in the initial dataset of Section~\ref{sec: Framework} are preserved. This is because without considering $\nu_x$, the flavor redistribution formula [Eq.~\eqref{eq : FE}] simply rescales both $\nu_e$ and $\bar\nu_e$ without flipping the sign of $G(\mathrm{v}_z,\phi_\nu)$.
The performance metrics are reported in Table~\ref{tab:FE modified maxent}. We see that the performance of our ML model on such a dataset is significantly high. This indicates that the model performs well as long as the input features of the test set are consistent with those used during training.  

Overall, these results highlight the importance of consistency between the input features of the training and test datasets for reliable model generalization. On the one hand, a physically consistent identification of the absence of crossings after flavor equilibration requires information from all neutrino species, including the heavy-lepton flavors. When such information is absent from the training inputs, the model performance deteriorates significantly. On the other hand, maintaining consistency between the feature spaces of the training and test datasets leads to a substantial improvement in performance.   
\begin{table}[tbh!]
\centering
\caption{The performance metrics of our models (trained in Section~\ref{sec: Framework}) tested on the dataset generated after applying flavor equilibration to the neutrino angular distributions. We note that the performance of the model is very poor as might be expected.}
\begin{tabular}[t]{|lcc|c|}
\hline
& \textcolor{black}{ \textbf{{Logistic Regression}} (53\%) }    \\
\hline
& precision & recall & $F_1$-score \\
\hline
no  crossing &100\%  & 53\% & 69\% \\
\hline
&\textcolor{black}{  \textbf{{KNN (k=7)}} (52\%) } \\
\hline
&precision&recall & $F_1$-score \\
\hline
no  crossing &100\%  & 52\% & 69\% \\
\hline
&\textcolor{black}{   \textbf{{SVM}} (51\%)}\\
\hline
&precision&recall & $F_1$-score\\
\hline
no  crossing &100\%  & 51\% & 67\% \\
\hline
&\textcolor{black}{   \textbf{{Decision tree}} (52\%)}\\
\hline
&precision&recall & $F_1$-score \\
\hline
no  crossing &100\%  & 52\% & 69\% \\
\hline
\end{tabular}
\label{tab:FE maxent}
\end{table}%
\begin{table}[tbh!]
\centering
\caption{Same as Table~\ref{tab:FE maxent}, but here the presence of crossings is determined only by the difference in $\nu_e$ and $\bar \nu_e$ distributions, i.e., the $\nu_x$, $\bar \nu_x$ differences are neglected. Note that all the models show a very good performance.}
\begin{tabular}[t]{|lcc|c|}
\hline
& \textcolor{black}{ \textbf{{Logistic Regression}} (93\%)}   \\
\hline
& precision & recall & $F_1$-score \\
\hline
no  crossing & 91\% &  96\% & 93\% \\
crossing&95\% & 91\% & 93\% \\
\hline
&\textcolor{black}{  \textbf{{KNN (k=7)}} (98\%) } \\
\hline
&precision&recall & $F_1$-score \\
\hline
no  crossing & 96\% & 100\% & 98\% \\
crossing&100\% & 96\% & 98\% \\
\hline
&\textcolor{black}{   \textbf{{SVM}} (98\%)}\\
\hline
&precision&recall & $F_1$-score\\
\hline
no  crossing & 98\% & 99\% & 98\% \\
crossing&99\% & 98\% & 98\% \\
\hline
&\textcolor{black}{   \textbf{{Decision tree}} (98\%)}\\
\hline
&precision&recall & $F_1$-score \\
\hline
no  crossing & 96\% &100\% & 98\% \\
crossing&100\% & 95\% & 97\% \\
\hline
\end{tabular}
\label{tab:FE modified maxent}
\end{table}

\section{Discussion and Conclusion}
\label{sec:Discussion}
In this work, we investigate the potential of ML methods to identify non-axisymmetric ELN crossings in CCSNe and binary NSM environments. The models are trained on artificially constructed maximum entropy angular distributions, with non-axisymmetry introduced through a relative frame rotation between the $\nu_e$ and $\bar \nu_e$ distributions. Their performance is evaluated on a diverse set of independent datasets, including realistic CCSN and NSM simulations as well as flavor-equilibrated scenarios.    

The models achieve relatively good predictive performance on both seen and most unseen datasets, demonstrating that the first two angular moments, including 
the azimuthal-angle-dependent information, capture the essential features associated with ELN crossings. Further, the variations in accuracy across different datasets indicate the influence of environment-dependent characteristics on model performance.  

Notably, for the 1D CCSN test set (Sec.~\ref{sec:Zewei's simulation})
our ML module's performance is mediocre for the original dataset that possesses axisymmetry, expanding the input features beyond  the original 
axisymmetric constraints to include non-axisymmetric neutrino angular information significantly 
enhances the 
performance.
This finding motivates further investigation into how the properties of training data and the choice of input features influence the generalization of ML models to completely unseen datasets.

A particularly interesting result emerges for flavor-equilibrated inputs. The model performs well when the training and test datasets contain the same flavor information, emphasizing the importance of consistency between the input feature spaces for reliable generalization. At the same time, our results show that a physically consistent determination of the absence of  $\nu$LN
crossings after flavor equilibration requires information from the heavy-lepton neutrino flavors. Since this information is absent from the input features used for training, the model cannot reliably identify all no crossing cases after equilibration.

In summary, our study demonstrates that ML provides a promising framework for identifying non-axisymmetric ELN crossings. The ability of the models to generalize from simplified, physically motivated training distributions to realistic simulation data suggests that the key characteristics governing ELN crossings can be effectively captured within an interpretable parameter space. These results represent a step toward integrating data-driven methods with state-of-the-art simulations of neutrino flavor evolution in CCSNe and NSMs.

\section*{Acknowledgments}
We are grateful to Manu George for the help in generating the NSM ray-tracing data.
We also thank Jakob Ehring and Georg Raffelt for useful discussions. 
MC and MRW acknowledge support from the National Science and Technology Council, Taiwan under Grant No.~111-2628-M-001-003-MY4, and Academia Sinica under Project No.~AS-IV-114-M04. 
MRW also acknowledges support from the Physics Division of the National Center for Theoretical Sciences, Taiwan. 
SA was supported by the German Research Foundation (DFG) through the Collaborative Research Centre ``Neutrinos and Dark Matter in Astro- and Particle Physics (NDM),'' Grant No.\ SFB-1258\,--\,283604770, and under Germany’s Excellence Strategy through the Cluster of Excellence ORIGINS EXC-2094-390783311.
FF acknowledges support from the Department of Energy, Office of Science, Office of Nuclear Physics, under contract number DE-SC0020435 and from the National Science Foundation through grant 2510568.
ZX acknowledges support of the European Research Council (ERC) through grant NeuTrAE (No.~101165138). The work is partially funded by the European Union. Views and opinions expressed are however those of the authors only and do not necessarily reflect those of the European Union or the European Research Council Executive Agency. Neither the European Union nor the granting authority can be held responsible for them. We would also like to acknowledge the use of the following softwares: \textsc{Scikit-learn}~\cite{pedregosa2011scikit}, \textsc{Matplotlib}~\cite{Matplotlib}, \textsc{Numpy}~\cite{Numpy}, and \textsc{SciPy}~\cite{SciPy}.

\bibliography{biblio}

\end{document}